\documentclass[aps,prl,twocolumn,showpacs,superscriptaddress,groupedaddress]{revtex4}  
\usepackage{aas_macros} 
\usepackage{graphicx}   
\usepackage{dcolumn}    
\usepackage{bm}         
\usepackage{amssymb}    
\usepackage{color}      

\newcommand{\nggn}{$(n,\gamma)\rightleftarrows(\gamma,n)$}  

\begin{document}

\hspace{5.2in} \mbox{LA-UR-16-21461}

\title{The link between rare earth peak formation and the astrophysical site of the $r$ process}

\author{M. R. Mumpower}
\affiliation{Theory Division, Los Alamos National Lab, Los Alamos, NM 87544, USA}
\affiliation{Department of Physics, University of Notre Dame, Notre Dame, IN 46556, USA}

\author{G. C. McLaughlin}
\affiliation{Department of Physics, North Carolina State University, Raleigh, NC 27695, USA}

\author{R. Surman}
\affiliation{Department of Physics, University of Notre Dame, Notre Dame, IN 46556, USA}

\author{A. W. Steiner}
\affiliation{Department of Physics and Astronomy, University of Tennessee, Knoxville, TN 37996, USA}
\affiliation{Physics Division, Oak Ridge National Laboratory, Oak Ridge, TN 37831, USA}

\date{\today}

\begin{abstract}
The primary astrophysical source of the rare earth elements is the rapid neutron capture process ($r$ process). 
The rare earth peak that is seen in the solar $r$-process residuals has been proposed to originate as a pile-up of nuclei during the end of the $r$ process. 
We introduce a new method utilizing Monte Carlo studies of nuclear masses in the rare earth region, that includes self-consistently adjusting $\beta$-decay rates and neutron capture rates, to find the mass surfaces necessary for the formation of the rare earth peak. 
We demonstrate our method with two types of astrophysical scenarios, one corresponding conditions typical of core-collapse supernova winds and one corresponding to conditions typical of the ejection of the material from the tidal tails of neutron star mergers. 
In each type of astrophysical conditions, this method successfully locates a region of enhanced stability in the mass surface that is responsible for the rare earth peak. 
For each scenario, we find that the change in the mass surface has qualitatively different features, thus future measurements can shed light on the type of environment in which the $r$ process occurred. 
\end{abstract}

\maketitle

\begin{figure}
 \begin{center}
  \centerline{\includegraphics[width=95mm]{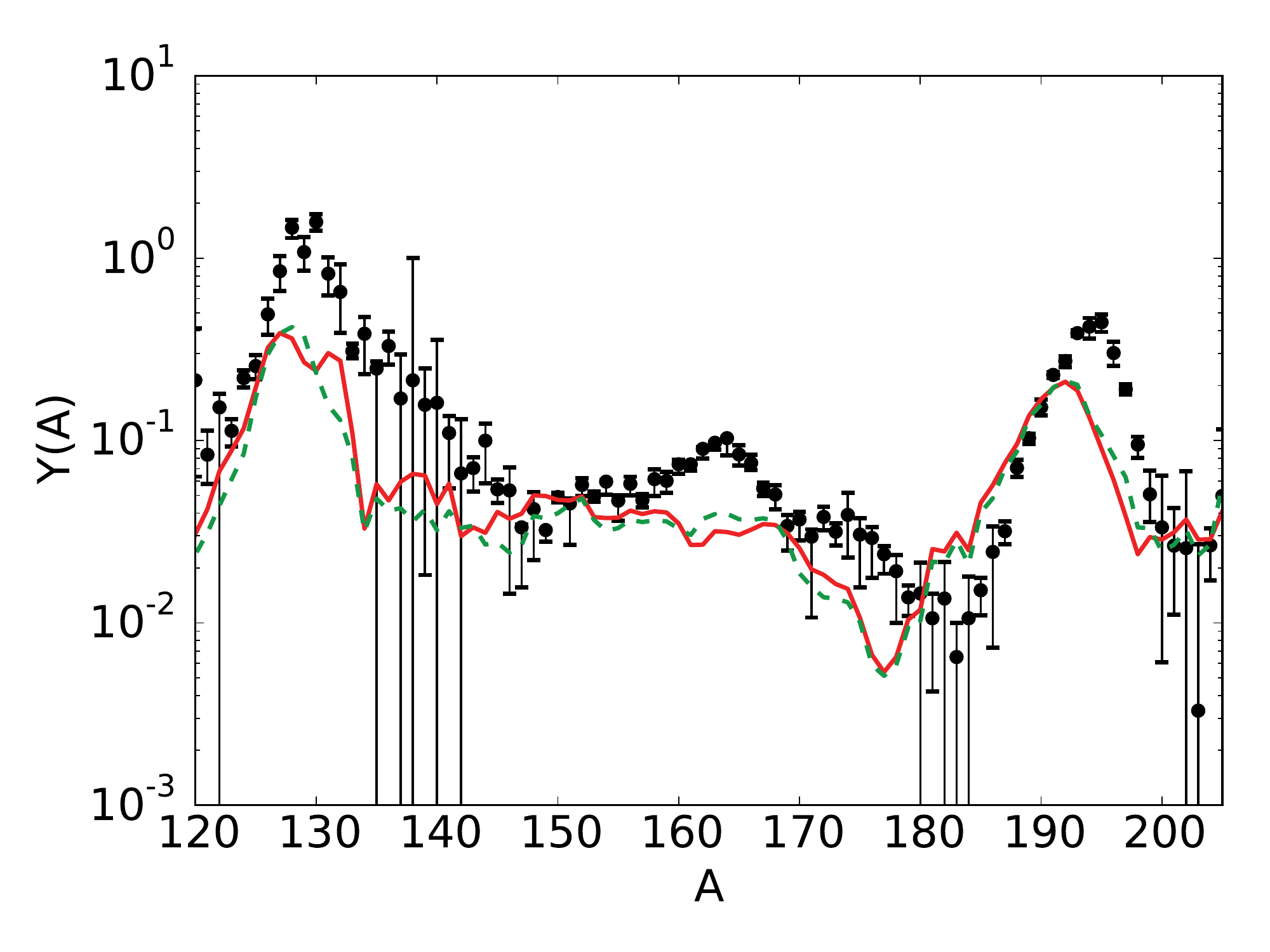}}
  \caption{\label{fig:ab} Simulations of the $r$ process with no rare earth peak in hot (red solid line) and very neutron-rich cold (green dashed line) conditions compared to the solar $r$-process residuals from Ref. \cite{Arnould+07} (black points).}
 \end{center}
\end{figure}

The majority of the solar system rare earth elemental abundances are attributed to the rapid neutron capture process of nucleosynthesis ($r$ process). 
While the basic mechanism of the $r$ process has been understood for some time \cite{Burbidge+57,Cameron+57}, there is recent evidence that $r$-process nuclei are formed in at least two separate ways \cite{Wasserburg+96,Sneden+08,Roederer+14,Hansen+14,Shibagaki+16}, sometimes called a weak $r$ process and a main $r$ process. 
The main $r$ process is what forms the rare earth elements, creating the peak at around $A=165$ shown in Fig.~\ref{fig:ab}. 
The astrophysical location of the main $r$ process has remained a mystery, despite numerous proposed sites; see reviews \cite{Qian+07,Arnould+07,Thielemann+11,Mumpower+16} and references therein. 
These sites include, among others, various locations within core-collapse supernovae (SN) and compact object mergers. 
The two sites that have received the most attention are the neutrino driven wind of core-collapse supernovae and ejection from the tidal tails of neutron star mergers. 

The ejecta from the tidal tails of neutron star mergers is a favorable main $r$-process site because it is a very neutron-rich environment and therefore guaranteed to make the even the heaviest $r$-process elements \cite{Lattimer+74,Meyer89}.  
In addition, it is considered a ``robust'' environment because it tends to produce a very similar pattern of abundances from the second peak up until the actinides \cite{Goriely+11,Korobkin+12,Wanajo+14,Rosswog+14,Just+15,Mendoza-Temis+15,Eichler+15}. 
This similarity in the pattern is observed in metal poor halo stars \cite{Sneden+08,Ji+15,Roederer+16,Sigueria-Mello+16}. 
Whether or not galactic chemical evolution studies can correctly predict the degree of scatter between stars in the overall level of $r$-process elements is currently under study \cite{Komiya+14,Matteucci+15,Wehmeyer+15,Shen+15,vandeVoort+15,Hirai+15}. 

Core-collapse supernovae are considered a favorable site for the $r$ process because galactic chemical evolution simulations find it easier to reproduce the scatter seen in the overall level of $r$-process material seen in old stars. 
The neutrino driven wind environment is close to producing the requisite neutrons for a complete main $r$ process \cite{Meyer+92,Woosley+94}, but so far self-consistent models \cite{Arcones+07,Fischer+10,Hudepohl+10,Roberts+12,Martinez-Pinedo+14}, in the absence of additional physics such as sterile neutrino oscillations \cite{McLaughlin+99}, do not quite make the entire pattern. 

The features seen in the $r$-process pattern arise from the interplay of nuclear masses, $\beta$-decay rates and neutron capture rates with the astrophysical temperature and density conditions as well as their gradients. 
The primary proposed mechanism by which the rare earth peak forms is through a dynamical process involving neutron capture at the late stages of the $r$ process when the nuclei decay back to stability \cite{Surman1997,Mumpower2012a}. 
In this scenario, a feature exists in the neutron separation energies or neutron capture rates which ``hangs up'' the nuclei in the rare earth region. 
Other places in the $r$ process where material becomes ``hung-up'' occur at closed neutron shells and correspond to the main $r$-process peaks (the second main peak at $A=130$ and the third main peak at $A=195$ can be seen in Fig.~\ref{fig:ab}). 
In these cases, a nuclear structure feature exists in stable nuclei and is assumed to extend into the region off of stability. 
In contrast, in the rare earth region no such feature is seen in the stable nuclei.
In order to confirm the theory of the dynamical formation mechanism, one needs to experimentally examine nuclei which are approximately ten to fifteen units in neutron number away from stability \cite{Mumpower2012c}. 

Some of the theoretical mass models commonly used in $r$-process calculations predict a nuclear physics feature away from stability that leads to dynamical rare earth peak formation, e.g.~\cite{FRDM1995}, though the peak is not always of the correct size and shape to match the solar pattern. 
Other mass models, e.g.~\cite{Goriely+13b}, show no such feature. 
Carefully-chosen linear combinations of astrophysical conditions have been shown to improve the fit to observation \cite{Kratz+14,Lorusso+15}. 
An alternate formation mechanism has been proposed that suggests the rare earth peak is made up of fission fragments resulting from a vigorous fission recycling $r$ process \cite{Schramm+71}. 
This mechanism hinges upon a specific distribution of fission daughter products \cite{Goriely+13} that is untestable by experiment. 
Thus, it can only be supported by indirect evidence, including the elimination of the dynamical mechanism as a viable alternative. 

In this letter, we introduce a new method by which the nuclear structure features that are necessary to produce characteristics of the $r$-process abundance pattern are determined by a Monte Carlo analysis. 
We apply this procedure to the portion of the isotopic solar abundances that includes the rare earth region, and we search for a persistent, non-local feature in the mass surface that leads to dynamical rare earth peak formation matching the solar pattern. 

There are two generic types of thermodynamic conditions that could exist toward the end of the $r$ process. 
We define ``hot'' environments as those where the material stays in \nggn \ equilibrium until the neutron number is no longer sufficiently high to maintain this equilibrium and ``cold'' environments as those where the equilibrium is broken because the temperature becomes too low. 
A standard supernova neutrino wind is a hot environment whereas the ejection of material from the tidal tails of neutron star mergers is both cold and very neutron rich. 
We apply our Monte Carlo procedure to both types of environments. 

As few mass measurements currently exist in the region in which we are interested, we require a theoretical baseline mass model. 
For our baseline model, we choose Duflo-Zuker (DZ) \cite{DZ} since it has little structure in the masses away from stability in the rare earth region. 
To verify this, we use the DZ mass model to compute neutron capture and beta decay rates and then run a set of $r$-process simulations for different astrophysical conditions. 
The neutron capture rates are computed using the Hauser-Feshbach code CoH \cite{Kawano2008}. 
For the $\beta$-decay rates, we use the underlying Gamow-Teller $\beta$-decay strength function, i.e.\ the nuclear matrix element information, from \cite{Moller2003}. 
We compute the phase space factor to be consistent with the DZ masses, as in Ref.~\cite{Mumpower2015b}. 
Our treatment of fission is largely schematic, as in \cite{Beun+08}, with spontaneous fission set to occur for $A>240$ and a simple asymmetric split assumed for the fission daughter product distributions.
This allows us to explore scenarios with fission recycling where the fission fragments ($A\sim130$) do not contribute to rare earth peak formation.
Examples of the results of $r$-process simulations with this set of nuclear data are shown by the red and green curves in Fig. \ref{fig:ab} for a hot and a cold very neutron-rich scenario, respectively. 
As expected the abundance pattern shows no feature in the rare earth region. 
This suggests the DZ mass model is missing the ingredient that leads to dynamical rare earth peak formation. 

Since we have a baseline model without structure in the rare earth region we are free to determine the missing component of the mass model which is required to match the $r$-process residuals. 
Previous studies have suggested that a kink in the separation energies as a function of neutron number is required \cite{Surman1997,Mumpower2012a}, but we wish to start with as little preconceived notion as possible about what this structure should be. 
Therefore, instead of choosing a parameterized form for a kink structure, we let an additional mass term float freely in neutron number, $N$: 
\begin{equation}
M(Z,N) = M_{DZ}(Z,N) + a_N e^{-(Z-C)^2/2f}
\label{eq:masses}
\end{equation}
Here, $M(Z,N)$ is the new mass generated from the baseline DZ mass, $M_{DZ}(Z,N)$, where $Z$ and $N$ represent the number of protons and neutrons in the nucleus. 
The ${a_N}$ are coefficients, one for each set of isotones with neutron number, spanning the range from $95$ to $115$. 
For a given neutron number, $a_N$ controls the overall magnitude and sign of the change to the base model. 
The parameter $C$ controls the center of the strength in proton number, and $f$ sets the fall off the strength in $Z$. 
The latter we keep fixed at $f=40$ because we are looking for a persistent feature in the mass surface. 

We now proceed to determine the $a_N$s and $C$ using the Metropolis algorithm \cite{Metropolis1953}. 
In brief, our procedure works as follows. 
As our initial guess we take all the $a_N=0$ and $C=60$, since $60$ is roughly the center of the rare earth peak in proton number for a typical late-time $r$-process path. 
With each change in masses we calculate self-consistently neutron capture rates and $\beta$-decay rates as previously described and we perform an $r$-process simulation, comparing the output to the $r$-process residuals. 
We then choose new values for the $a_N$s and $C$ and repeat. 
At each step the parameters $a_N$ are chosen from a normal distribution centered at the current value with a spread of $0.025$ MeV, and the parameter $C$ is chosen from a normal distribution with spread $0.1$. 
The current values of the parameters are updated in accordance with the Metropolis prescription, which seeks to optimize the output of the reaction network to the observed $r$-process residuals.
We find our calculations converge in approximately 20,000 to 30,000 steps for each astrophysical trajectory considered.

\begin{figure}
 \begin{center}
  \centerline{\includegraphics[width=95mm]{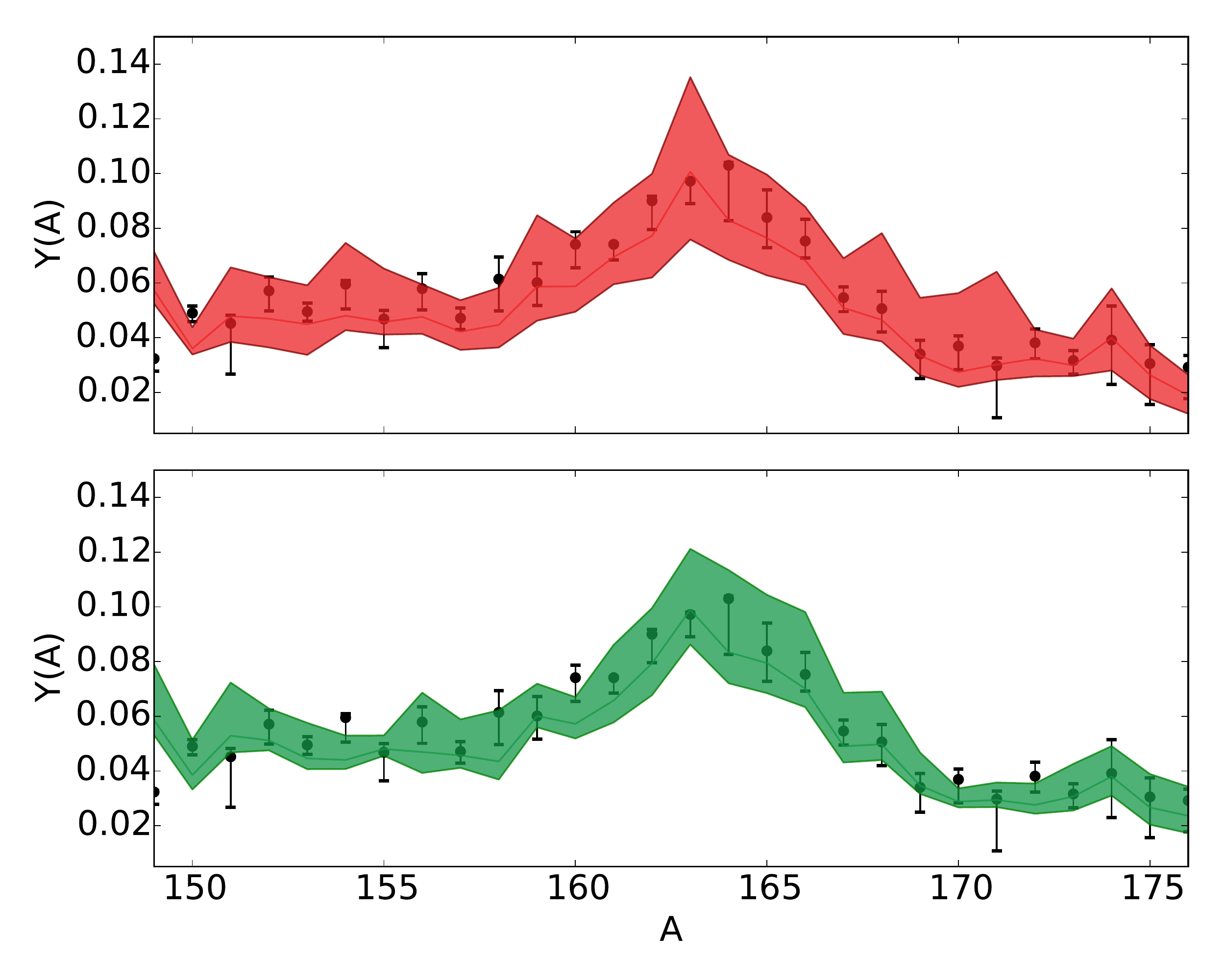}}
  \caption{\label{fig:delta} The final rare earth peak abundances for hot (top panel) and very neutron-rich cold (bottom panel) $r$-process conditions, compared to solar $r$-process residuals from Ref. \cite{Arnould+07} (black points). The final abundance uncertainties, denoted by the shaded regions, originate from the predictions in the mass surface after application of the Metropolis algorithm. }
 \end{center}
\end{figure}

In Fig.~\ref{fig:delta} we display the calculated final $r$-process abundances using the new predictions of nuclear masses in the rare earth region from the application of our framework. 
To construct the shaded bands we compute the averages and standard deviations for multiple Metropolis algorithm runs, with the averaging performed separately for hot and very neutron-rich cold trajectories. 
This ensures we have sufficient statistics and that we draw general conclusions which are not based on particular details of a single trajectory. 
The hot conditions used are parameterized winds all with long duration \nggn \ equilibrium having entropies $30$, $200$, and $110$ in units of $k_{B}$/baryon with timescales $70$, $80$, and $160$ in units of ms and electron fractions $0.2$, $0.3$, and $0.2$, respectively \cite{Mumpower2012b}. 
The very neutron-rich cold conditions used are from Refs.~\cite{Goriely+11} and \cite{Just+15}. 
From Fig.~\ref{fig:delta} we see that this algorithm produces an excellent match to the $r$-process residuals. 
Both the overall and the subtle features of the pattern are reproduced in both hot (top panel) and very neutron-rich cold scenarios (bottom panel). 

\begin{figure}
 \begin{center}
  \centerline{\includegraphics[width=95mm]{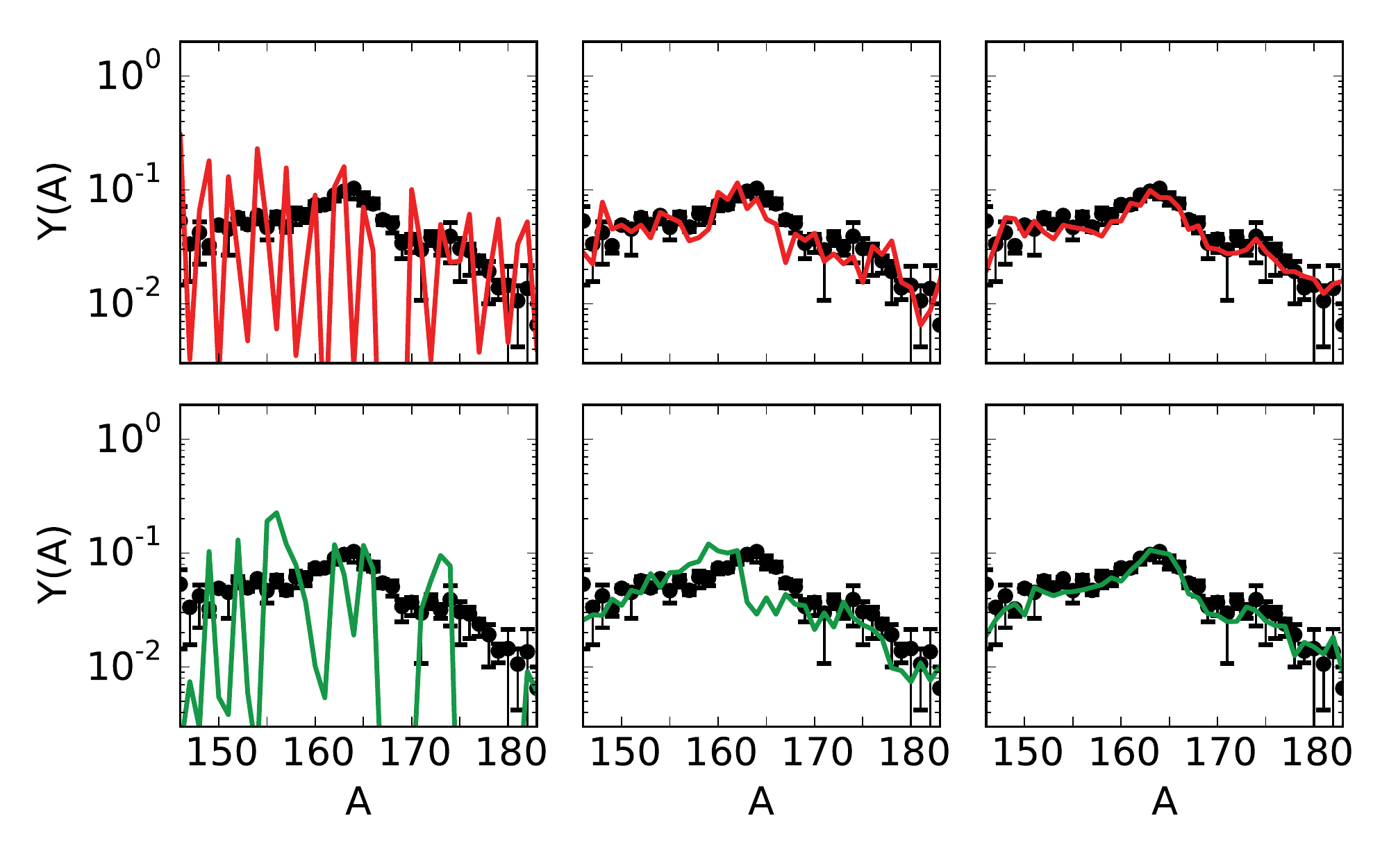}}
  \caption{\label{fig:snapshots} Evolution and formation of the rare earth peak via the dynamical neutron capture mechanism in a hot (top panel) and a very neutron-rich cold (bottom panel) $r$ process. Snapshots show the break from \nggn \ equilibrium (left), the start of peak formation (center), and the final abundances (right).}
 \end{center}
\end{figure}

We use the results of our calculations to examine more closely the dynamical formation mechanism of the rare earth peak predicted by the algorithm for hot and very neutron-rich cold scenarios. 
Three stages in the evolution of the rare earth peak are depicted for both types of scenarios in Fig.~\ref{fig:snapshots}, hot in the top set of panels and very neutron-rich cold in the bottom set. 
In the hot calculation there is little sign of the rare earth peak during the majority of the $r$ process (left panel), but as the rare earth peak begins to form during the late stages (middle panel) it forms in the same mass number region as the peak in the $r$-process residuals. 
In contrast, when the rare earth peak first forms in the cold scenario (bottom middle panel), it forms at slightly lower mass number than the $r$-process residuals and it is late time neutron capture that moves the peak to its correct position (bottom right panel). 

\begin{figure}
 \begin{center}
  \centerline{\includegraphics[width=95mm]{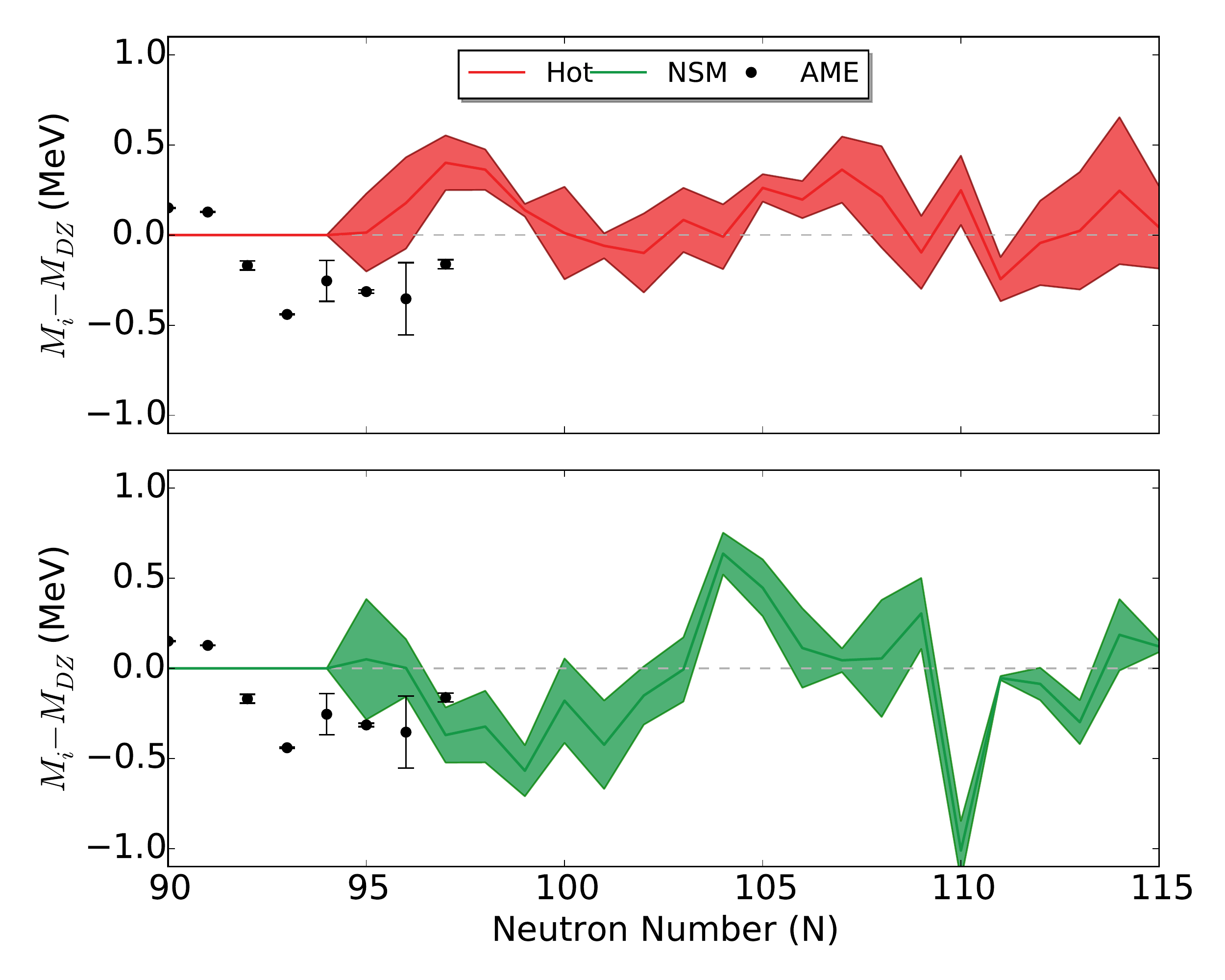}}
  \caption{\label{fig:isochain} Differences in mass datasets from Duflo-Zuker along the $Z=60$ (Nd) isotopic chain. The shaded regions show the predicted change to the Duflo-Zuker mass surface using the Monte Carlo technique from this paper for a hot (red) and very neutron-rich cold (green) $r$-process. Points show experimental data from the latest Atomic Mass Evaluation \cite{AME2012}. }
 \end{center}
\end{figure}

Because we observe that the rare earth peak forms differently in hot and very neutron-rich cold scenarios as shown in Fig.~\ref{fig:snapshots}, we anticipate that the required structure in the mass surface is different for different astrophysical conditions. 
We investigate this by looking at the modifications to the DZ masses which correspond to Fig.~\ref{fig:delta}. 
These are shown by shaded bands in Fig.~\ref{fig:isochain}. 

In the top panel of Fig.~\ref{fig:isochain} we show the result for hot scenarios and in the bottom panel we show the result for very neutron-rich cold scenarios along the $Z=60$ (Neodymium) isotopic chain. 
The most striking feature of this figure is that in both types of astrophysical conditions a dip in this curve in the general region of $N=100$ is required. 
The dip represents a region that has enhanced stability, allowing material to be hung up when the $r$-process path passes through it.
Thus, this feature is primarily responsible for the formation of the rare earth peak. 

A second important observation from this figure is that the shape of these two curves is different for the two distinct sets of astrophysical conditions, both in the depth of the dip and in its location. 
Under hot $r$-process conditions the relative minimum is relatively shallow, from highest to lowest points spans no more than $1$ MeV. 
The shape of the curve under very neutron-rich cold conditions shows a larger range in the mass differences with a span of over $1$ MeV between the highest and lowest points. 

The position of the local minimums relative to the DZ masses also differs. 
For hot conditions the minimum is around $N=100$, $102$, and $104$. 
For cold scenarios, it is shifted to lower $N$, consistent with an initial formation of the peak at lower mass number $A$. 
In these scenarios, the minimum of the mass surface relative to DZ is around $N=97$, $99$ and $101$. 
This propensity of the system to favor even neutron numbers in hot scenarios and odd neutron numbers in cold scenarios is connected to the formation mechanism. 
For hot conditions, in which \nggn \ equilibrium persists for long times, the material tends to collect in even masses \cite{Surman1997}. 
But for cold scenarios, neutron capture rates are most important, thus favoring odd masses \cite{Mumpower2012a,Mumpower2012c}. 
A second strong feature is seen at $N=110$ for very neutron-rich cold scenarios. 
We find this feature to be required to fill in the hole to the right of the rare earth peak that exists in our baseline model, shown in Fig.~\ref{fig:ab}. 

The mass surface for the Neodymium isotopes are highlighted in Fig.~\ref{fig:isochain}. 
We set out to find a global feature in the masses, slowly varying in $Z$, and so we kept the falloff parameter $f$ at a fixed, large value. 
We allow the center in $Z$, $C$, to float, and we find that as long as the initial value is around $C\sim 58-62$ it does not vary much upon application of the Metropolis algorithm. 
The modeled mass surface changes therefore are similar for all the isotopic chains surrounding Neodymium. 
Our tests with smaller fixed values of $f$ show that additional solutions may be possible for a feature more tightly localized in $Z$, however such localized features need to be significantly larger than those shown in Fig.~\ref{fig:isochain} for the simulations to produce a good match to the solar pattern. 

Since there is a clear difference between the predicted mass surface for neutron star mergers and core-collapse supernovae, future mass measurements in the rare earth region can shed considerable light on the astrophysical scenario of the main $r$ process by determining whether there is a region of enhanced stability in the mass surface and, if so, its depth and position. 
Dips at lower neutron number or local minimums at odd nuclei would favor mergers, whereas dips at high neutron number or local minimums at even nuclei would favor hot scenarios such as supernova winds. 
Our framework is sufficiently general such that the favored mass surface for robust rare earth peak formation in other proposed sites of the $r$ process can be analyzed in a similar manner.

If no discernible shape in the mass surface akin to that shown in Fig.~\ref{fig:isochain} is found, then either the rare earth peak forms dynamically in a way not captured by our model---the site of the $r$ process is one we have not considered, e.g.~\cite{Surman+08,Metzger+08,Barnes+13,Banerjee+11,Winteler+12,Nakamura+15}, or the mass feature responsible is a sharp and local instead of smooth and global---or the rare earth peak comes from the daughter products of fission cycling. 
The latter outcome strongly favors compact object mergers as the site of the $r$ process and would have implications for our understanding of the fission properties of heavy neutron-rich nuclei.

The recent advent of Penning trap and ion storage ring technology has spurred a significant increase in both the quantity and quality (high precision) of nuclear mass measurements \cite{Lunney+03,Blaum+06,Baruah+08,Hakala+12,VanSchelt+13,Sun+15,Atanasov+15,Klawitter+15,Lascar+15}. 
The coupling of these techniques with radioactive ion beam facilities and future technological advances will open extensive regions of the nuclear chart for measurement \cite{Lorusso+15,Kurtukian+14,Spyrou+14,Cizewski+15,Domingo+16}, including neutron-rich nuclei of interest for the $r$ process \cite{Mumpower+16}. 
It is not unreasonable to expect that in the next few years, new measurements will be able to shed light on what sort of astrophysical conditions the rare earth peak, and therefore the main $r$ process, formed. 

\acknowledgments
This work was supported in part by the National Science Foundation through the Joint Institute for Nuclear Astrophysics grant numbers PHY0822648 and PHY1419765 (MM), and the Department of Energy under contract DE-SC0013039 (RS) and  U.S. DOE Grant No. DE-FG02-02ER41216 (GCM). 
A portion of this work was also carried out under the auspices of the National Nuclear Security Administration of the U.S. Department of Energy at Los Alamos National Laboratory under Contract No. DE-AC52-06NA25396 (MM). 

\bibliographystyle{unsrt}
\bibliography{refs}

\end{document}